\documentclass{sigkddExp}

\begin{document}
%

\title{Sentiment Analysis: A Survey}
%

\numberofauthors{1}
%


\author{
%
\alignauthor Rahul Tejwani \\
       \affaddr{University at Buffalo}\\
       \affaddr{Buffalo, New York}\\
       \email{rahultej@buffalo.edu}
}

\date{15 April 2014}
\maketitle
\begin{abstract}
Sentiment analysis (also known as opinion mining) refers to the use of natural language 
processing, text analysis and computational linguistics to identify and extract subjective information in source materials.
Mining opinions expressed in the user generated content is a challenging yet practically very useful
 problem.

This survey would cover various approaches and methodology
used in Sentiment Analysis and Opinion Mining in general. The focus
would be on Internet text like, Product review, tweets and other social media.
\end{abstract}

\section{Introduction}
Humans are subjective creatures and
opinions are important. Being able to
interact with people on that level has many
advantages for information systems.
Comparatively few categories
(positive/negative, 3 stars, etc) compared
to text categorization
Crosses domains, topics, and users
Categories not independent (opposing or
regression-like)
Characteristics of answers to opinion-
based questions are different from fact-
based questions, so opinion-based information extraction
differs from trad information extraction.

Some of the challenges in Sentiment Analysis are:
People express opinions in complex ways,
in opinion texts, lexical content alone can
be misleading.
Another challenge can be in the form of Intra-textual and sub-sentential reversals,
negation, topic change common.
Humans tend to express a lot of remarks in the form of 
sarcasm, irony, implication, etc. which is very difficult
to interpret. For Example- ``How can someone sit through the movie'' is extremely negative
sentiment yet contains no negative lexographic word.
Even if a opinion word is present in the text,their can be cases where a opinion word that is considered to be 
positive in one situation may be considered negative in 
another situation. People can be contradictory in their statements. Most 
reviews will have both positive and negative comments, 
which is somewhat manageable by analyzing sentences 
one at a time. However, in the more informal medium 
like twitter or blogs(Social media), the more likely people are to 
combine different opinions in the same sentence which is 
easy for a human to understand, but more difficult for a 
computer to parse. Sometimes even other people have 
difficulty understanding what someone thought based on 
a short piece of text because it lacks context. A good example would
be ``The laptop is good but I would prefer, the operating system 
which I was using''.

There is a huge demand of sentiment analysis. Before buying
 any product its a practice now, to review its rating as rated by other
 persons who are using it. Online advice and recommendations the data reveals is 
 not the only reason behind the buzz in this area. There are other reasons like 
 the company wants to know ``How Successful was their last campaign or product launch'' based upon 
 the sentiments of the customers on social media.
.Sentiment analysis concentrates on attitudes, 
whereas traditional text mining focuses on the analysis of 
facts. There are few main fields of research predominate 
in Sentiment analysis: sentiment classification, feature 
based Sentiment classification and opinion 
summarization.

Main research in the area of Sentiment Analysis and opinion mining 
are: sentiment classification, feature 
based Sentiment classification and opinion 
summarization. Sentiment classification deals with 
classifying entire documents or text or review according to the opinions 
towards certain objects. Feature-based Sentiment 
classification on the other hand considers the opinions on 
features of certain objects. For example, in reviews related to laptops
classifying the sentiments only on the basis screen quality.
 The task of Opinion summarization is 
different from traditional text summarization because 
only the features of the product are mined on which the 
customers have expressed their opinions. Opinion 
summarization does not summarize the reviews by 
selecting a subset or rewrite some of the original 
sentences from the reviews to capture the main points as 
in the classic text summarization.

Languages that have been studied mostly are English 
and in Chinese .Presently, there are very few researches 
conducted on sentiment classification for other languages 
like Arabic, Spanish, Italian and Thai. This survey aims to focus 
on English only.

\section{Task Sub-Division}

The task of Sentimant Analysis can be broadly classified into
 three levels:
document level, 
sentence level,
and feature based approaches (aspect level).[Bing Liu]

\subsection{Document level Sentimant Analysis}
classification of the overall sentiment of a document
 based on the overall sentiment of the
 opinion holder. This problem is basically
 a text classification problem. Here in general it is assumed
 that the document is written by a single person and
 expresses opinion about a single entity. One of the major challenge in the 
document level classification is that all the sentence in a
document may not be relevant in expressing the opinion 
about an entity. Therefore subjectivity/objectivity 
classification is very important in this type of classification. 
The irrelevant sentences must be eliminated from the 
processing works.
 
 Both \textbf{supervised and unsupervised learning} methods can be 
used for the document level classification. Any supervised 
learning algorithm like naïve Bayes classifier, Support Vector 
Machine, or Maxmimum Entropy etc can be used to train the system. For training and 
testing data, the reviewer rating (in the form of 1-5 stars), 
can be used. The features that can be used for the machine 
learning are term frequency, adjectives from Part of speech 
tagging, Opinion words and phrases, negations, 
dependencies etc. Labeling the polarities of the document 
manually is time consuming and hence the user rating 
available can be made use of. The unsupervised learning can 
be done by extracting the opinion words inside a document. 
The point-wise mutual information can be made use of to 
find the semantics of the extracted words. Thus the 
document level sentiment classification has its own 
advantages and disadvantages. Advantage is that we get an 
overall polarity of opinion text about a particular entity from 
a document. Disadvantage is that the different emotions 
about different features of an entity could not be extracted 
separately.

\subsection{Sentence level Sentimant Analysis}
Here polarity of each contributing senetence is derived. Again, here the 
assumption is that each sentence is written by a
single person and expresses a single positive
or negative opinion/sentiment.
Sometimes Document-level sentiment classification is too coarse
for our pupose. One of the reason can be, the size of the
document is too large or a more granular level of sentiments needs
 to be derived. A lot of early work in the region ofsentence level analysis
focuses on identifying subjective sentences. Most techniques use supervised learning.

This can be devided into two task: first identify which sentence old opinion (subjective sentences) and 
then classify each sentence as positive/negative or the star rating. But 
there will be complex sentences also in the opinionated text. 
In such cases, sentence level sentiment classification is not 
desirable. Knowing that a sentence is positive or negative is 
of lesser use than knowing the polarity of a particular 
feature of a product. The advantage of sentence level 
analysis lies in the subjectivity/ objectivity classification.

Some challenges in this approach could be:
many objective sentences can imply sentiments
 or Many subjective sentences do not express
positive or negative sentiments/opinions. A single sentence may contain multiple
opinions and subjective and factual clauses.

\subsection{Feature based or Aspect level Sentiment Analysis}
A more granular approach that gives some extra information. For example``
Sentiment classification at both the document
and sentence (or clause) levels are useful, but they do not find what people liked and disliked. 
The product or the review.They do not identify the targets of opinions. Much of the research is based on online reviews and
blog related data.
In the cae of reviews, where the entity (product or service) is known. Its a easier 
problem.
But for blogs, forum discussions, etc., it much harder because
the entity is unknown there may also be many comparisons, and
there is also a lot of irrelevant information. This problem is 
somewhat similar to the problem of Named Entity Resolution.

Another interesting approach to solve sentiment analysis
was presented in [4] by exploiting concept chains to build a graph 
model of a text. The model used Part of speech tagging to extract
graph relations and performed graph based queries to get the overall
sentiment of the text.

\section{Data}
Major forms of data available for Sentiment Analysis are:
Blogs, review sites, data 
and lexical dictionaries.
\subsection{Data Sets availble}
Most of the work in the field uses movie 
reviews data for classification. Movie review datas are 
available as dataset 
{http://www.cs.cornell.edu/People/pabo/movie-review-data}. 
Other dataset which is available online is multi-domain 
sentiment (MDS) dataset. 

{http://www.cs.jhu.edu/mdredze/datasets/sentiment}. The MDS 
dataset contains four different types of product reviews 
extracted from Amazon.com including Books, DVDs,.
Electronics and Kitchen appliances, with 1000 positive 
and 1000 negative reviews for each domain. Another 
review dataset available is 

{http://www.cs.uic.edu/liub/FBS/CustomerReviewData.zip}. 
Multi-Perspective Question Answering
(MPQA) (Stoyanov et al, 2005): News articles and other text documents
manually annotated for opinions and other
private states (i.e., beliefs, emotions,
sentiments, speculations, etc.)consists of 
692 documents (15,802 sentences)
(http://www.cs.pitt.edu/mpqa/)

\subsection{Review Sites}
These days almost every e-commerce website 
have reviews for their product. Some of the famous review sites which
have made their data available for research are: www.amazon.com 
(Product review), www.yelp.com (restaurant reviews), 
www.CNET download.com (product reviews) and 
www.reviewcentre.com, which hosts millions of product 
reviews by consumers.

\subsection{Lexical Dictionaries}
There are many dictionaries available that have information about, 
opinion words and polarity. Some of the famous lexicon are:
Sentiwordnet 3.0 ({http://sentiwordnet.isti.cnr.it/}),it
assigns to each synset of WordNet three sentiment scores: positivity, negativity, objectivity 
LIWC: (http://www.liwc.net/), Linguistic Inquiry and Word Count
General Inquirer : 

(http://www.wjh.harvard.edu/~inquirer/)
: Database of words and manually created
semantic and cognitive categories,
including positive and negative
connotations

\section{subjectivity and objectivity Classification}
To determine whether a text holds any Subjective information or
opinion Subjectivity analysis is performed. 
The text pieces may or may not contain useful opinions or 
comments. The subjective sentences are the relevant texts, 
and the objective sentences are the irrelevant texts. So we 
must sort out the sentences that are useful for us and those 
which are not. The subjective sentences are those sentences 
having useful information for the sentiment analysis. Such 
classification is termed as subjectivity classification. Some 
works have been done focusing on this particular problem. 
 
In one of the most poineer work [1], 
the authors present a method of subjectivity 
identification for sentiment analysis. This is important 
because the irrelevant data from the reviews could be 

eliminated. This eliminates the processing overheads of a 
large amount of textual data. The method they propose is 
using minimum cuts to produce subjective extracts from the 
text. The work has been focused in the sentence level 
subjectivity extraction

\section{Applications}
Sentiment analysis has been a buzz recently
and a lot of research is been going on in this area.
These days it used in social media monitoring
 and VOC(voice of customer) to track customer reviews, survey responses, competitors, etc. 
However, it is also practical for use in business analytics
 and situations in which text needs to be analyzed. 
 
Sentiment analysis is in demand because of its efficiency.
 Thousands of text documents can be processed for sentiment
 (and other features including named entities, topics, themes, etc.) in seconds,
 compared to the hours it would take a team of people to manually complete. 
Because it is so efficient (State of the art being around 80 percent )
 many businesses are adopting text and sentiment analysis 
and incorporating it into their processes.

Sentiment analysis also have a lot of
applications in Business intelligence, since it is very difficult
and expensive process to survey customers. It also
answers some of questions such as:
 “Why is the product not selling?”
  “What are the specific issues faced by users?”
\section{Conclusion}
There are various methods to classify text. Several techniques 
involves word and phrase classification methods exist. Many Lexical and
dictionary based approach [11] are used and have been fairly Successful 
some of them even showed more than 70 percent accuracy on standard data sets.

Another approach that is more commonly used is Machine Learning. Various Machine 
learning techniques have been tried and used in the papers mentioned. The most common ones are: Naive Bayesian,
Support Vector Machine[9] and Maximum entropy. The success of a  machine learning algorithm depends
upon how well the features are selected and extracted. Various different features have been used, the ones that are most 
commomn are Unigrams, Parts of speech tagging, emoticons detection, categories tagging,
using graph based techniques. Topic modelling [10] and phases based features [8] are also used.

%
\bibliographystyle{abbrv}
.
\bibliography{sigproc}  
%
%

[1] B. Pang and L. Lee, “A sentimental education: 
Sentiment analysis using subjectivity summarization based 
on minimum cuts,” 

[2] Bo Pang , Lillian Lee and Shivakumar Vaithyanathan
''Thumbs up? Sentiment Classification using Machine Learning Techniques``

[3] A Pak, P Paroubek, ''Twitter as a Corpus for 
Sentiment Analysis and Opinion Mining:`` 

[4] Wei jin, Srihari, ''Mining Hidden Associations
in Text Corpora through
Concept Chain and Graph Queries``

[5] Bing Liu, ''Sentiment Analysis and Subjectivity`` 

[6] Minqing Hu and Bing Liu, ''Opinion Extraction
and Summarization on the Web`` 

[7] Bai, and R. Padman, “Markov blankets and meta-heuristic 

search: Sentiment extraction from unstructured text,”

[8] Theresa Wilson, Paul Hoffmann, ''An Exploration 
of Features for Phrase-Level Sentiment Analysis``

[9] Tony Mullen and Nigel Collier, ''Sentiment analysis

using support vector machines with diverse information sources`` 

[10] Chenghua Lin, Yulan He, ''Joint Sentiment/Topic Model
for Sentiment Analysis`` 

[11] Maite Taboada, Julian Brooke, Kimberly Voll, Manfred 
Stede, ''Lexicon-Based Methods for Sentiment Analysis``

\end{document}